\documentclass[12pt,endfloat]{article}
\usepackage{varioref}
\usepackage{graphicx}
\usepackage[T1]{fontenc}
\usepackage{epsfig}
\usepackage{subfigure}
\usepackage{cite}
\usepackage[labelsep=space]{caption}
\usepackage{amsmath}
\usepackage{amsfonts}
\usepackage{stmaryrd}
\usepackage{psfrag}
\usepackage[a4paper]{geometry}
\usepackage{setspace}
\usepackage{color}
\usepackage{tabularx}\newcolumntype{Y}{>{\centering\arraybackslash}X}

\begin{document}
\title{Correcting the formalism governing Bloch Surface Waves excited by 3D Gaussian beams}
\author{Fadi I. Baida$^{*,1}$ \& Maria-Pilar Bernal$^{1}$\\
$^{1}$ D\'epartement d'Optique P.M. Duffieux,\\ Institut
FEMTO-ST, UMR CNRS 6174, \\Universit\'e  de Franche-Comt\'e, \\25030 Besanon cedex, France\\
$^*$ email: fbaida@univ-fcomte.fr }
\date{}

\maketitle  
\setstretch{1,1}

\begin{abstract}
Due to the growing number of publications and applications based on the exploitation of Bloch surface waves and the gross errors and approximations that are regularly used to evaluate the properties of this type of wave, we judge seriously important for successful interpretation and understanding of experiments to implement adapted formalism allowing to extract the relevant information. Through a comprehensive calculation supported by an analytical development, we establish a generalized formula for the propagation length which is different from what is usually employed in the literature. We also demonstrate that the Goos-H\"{a}nchen shift becomes an extrinsic property that depends on the beam dimension with an asymptotic behavior limiting its value to that of the propagation length. The proposed theoretical scheme allows predicting some new and unforeseen results such as the effect due to a slight deviation of the angle of incidence or of the beam-waist position with respect to the structure. This formalism can be used to describe any polarization-dependent resonant structure illuminated by a polarized Gaussian beam.  
\end{abstract}

 \newpage
\setstretch{1,4}

Quantification of the $BSW$ properties is very important to predict its effectiveness in being used in integrated or in surface optics. $BSW$s are electromagnetic surface modes used to design different configurations for applications ranging from sensing \cite{guillermain:apl07,petrova:sr19} to surface-optics \cite{sfez:apl10,wang:natcom17,kim:acsp17,lahijani:ol17,wang:acs17,wang:lsa18,augenstein:lsa18,kim:cp18} or micro-manipulation \cite{shilkin:ol15,shilkin:acs17}. This kind of surface waves is of a great interest in integrated optics \cite{descrovi:nl10,lu:lsa13} due its very large propagation distance and the possibility to be excited in both $TE$ and $TM$ polarizations contrarily to surface plasmon. Similarly to the latter, $BSW$ can either be excited in the Kretschmann configuration (total internal reflection) \cite{yeh:apl78,descrovi:apl07} or more simply by diffraction \cite{liscidini:apl07,kolvalevich:oex17}. However, 3D $BSW$ electromagnetic field distribution has never been theoretically reported except very recently by pure numerical methods (Finite Difference Time Domain \cite{wang:acs17} or Finite elements \cite{koju:sr17}). This is a prerequisite for evaluating the two most important properties of $BSW$, namely its propagation length $PL$ and lateral or Goos-H\"{a}nchen shift $L_{GH}$, which will be defined later on. This will be addressed through two different ways: (a) a rigorous method based on the Transfer Matrix Method $TMM$ combined to the description of a 3D polarized Gaussian beam by an accurate Plane Wave Expansion $PWE$ and, (b) an analytical calculation of the electromagnetic field associated to the $BSW$ itself. As it will be demonstrated, the two methods converge to the same result that fails the commonly used formulas.  For the $PL$, we establish a new equation that is widely valid for any surface wave excited within high quality-factor resonance having a Lorentzian shape (surface plasmon, Fano, membrane mode, symmetry protected modes, Bounded in the Continuum modes...). For the $L_{GH}$, we demonstrate its value to be dependent on the incident beam dimension, which is completely innovative, compared to the accepted ideas for which this property is intrinsic to the structure itself. 

As mentioned below, our findings are in great contradiction with commonly used formulas. On one hand, several studies \cite{descrovi:oe08,soboleva:prl12,angelini:ol14,dubey:eos17} used theoretical formulas based on a development obtained for plane wave illumination \cite{artmann:anph48,ulrish:josa71}. For example, in ref. \cite{soboleva:prl12}, the formula given in Eq. 1 of that paper is used to discuss the occurrence of a giant Goos-H\"{a}nchen shift on the reflected beam issued from the excitation of a $BSW$. In that paper, the measured reflectance angular spectrum is used to estimate the Fano profile of the resonance then operated to evaluate the lateral shift of the reflected beam. Nevertheless, as in most theoretical studies \cite{andaloro:josab05,simon:ol07,konopski:pra12}, the approach used to assess the reflected beam distribution is based on the consideration of one-dimensional angular distribution for the beam (see Eq. 3 in \cite{soboleva:prl12}) meaning that the incident beam is a 2D Gaussian beam (prismatic) instead of a realistic 3D beam. This certainly leads to less reliable physical properties of the studied phenomenon as it will be discussed in more details below. 
On the other hand, in diverse studies as in \cite{soboleva:prl12}, the use of the reflectance spectrum to estimate the $BSW$ properties is somewhat questionable. In fact, the $BSW$ corresponds to a surface mode that is excited in the total internal reflection condition meaning that the reflection coefficient is equal to $100\%$ in amplitude for purely dielectric flat layers. Consequently, the signature of the $BSW$ excitation on the reflection coefficient only involves its phase and never its amplitude nor its intensity that is usually experimentally measured. When the reflectance spectrum exhibits a dip resonance, this gives directly the effective index of the $BSW$ (through the tangential wave-vector component) but means above all that losses occur by scattering or by absorption. In this case, the relationship between the angular width of the reflectance dip and the $BSW$ properties is no longer intuitive.  

\underline{\textbf{Proposed structure and plane wave analysis}}\newline First, we consider a typical configuration of 1D-PhC by optimizing its geometry using a plane wave illumination through a very simple algorithm based on $TMM$ (see details in \textbf{SI} file) that links the electric incident and reflected field amplitudes to the transmitted and back reflected ones on the interface separating two different layers. The total transfer matrix, which is the product of all single matrices, allows determining the transmitted and reflected amplitudes over the entire multi-layered system as a function of the incident one (see Eqs E.4 and E.5 of the \textbf{SI} file) by taken into account all the geometrical and physical parameters of the structure (thicknesses and permittivities) and the plane wave properties (polarization, wavelength, angle of incidence). The eigenvalues of this total matrix are the eigenmodes of the structure that can be simply calculated through basic inverse matrix algorithm.

\begin{figure}
\centering
\includegraphics[width=14cm]{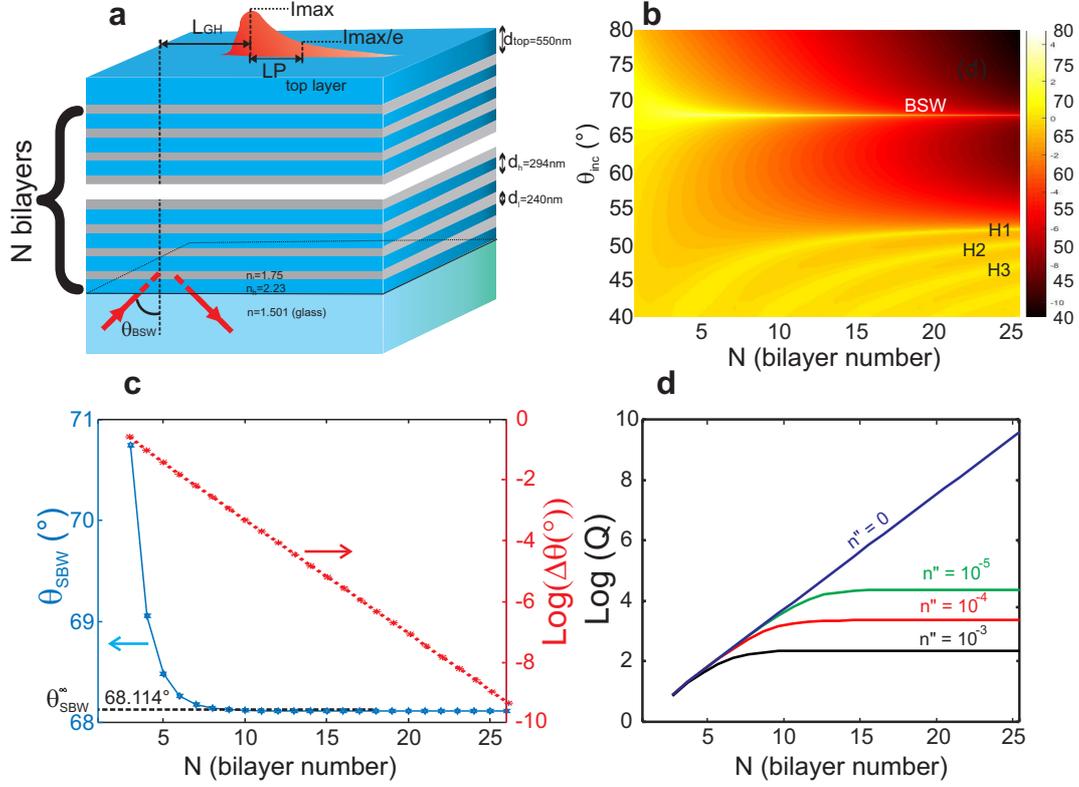}
\caption{\textbf{| Schematic of the studied 1D-PhC structure and transmission properties. a,} The incident beam illuminates the structure from a glass substrate at an angle $\theta_m$. It is linearly polarized along the $y-$direction ($TE$ polarization). The bi-layer stack is composed of a layer of high index media ($n_h=2.23$) with a thickness $d_h=294~nm$ deposited on a second layer made in low index material ($n_l=1.75$) of thickness $d_l=240~nm$. The total number of stacks is named $N$ and the structure is terminated by a top layer of high index material of $d_{top}=550~nm$ thickness. \textbf{b,} Transmitted electric intensity in logarithmic scale versus the number of bi-layers $N$. Additional modes occur when $N$ increases corresponding to smaller values of the angle of incidence. \textbf{c,} Variations of $\theta_{BSW}$ as a function of the number of bi-layers ($N$) in blue solid line and its $FWHM$ variation $\Delta\theta$ (in log-scale) in red dashed line. \textbf{d,} Variations of the quality factor $Q$ of the $BSW$ excitation as a function of the bilayer number $N$ for different imaginary part $n''$ values.}
\label{tetaNN}
\end{figure}

We use this kind of calculation to adapt a multilayer design \cite{descrovi:oe08,sfez:josab10} that consists on $N$-periods of bi-layered stacks (see Fig. \ref{tetaNN}a) to operate at telecoms wavelength in $TE$ polarization. All geometrical parameters are given in the caption the figure \ref{tetaNN}. 

Figure \ref{tetaNN}b shows the square modulus of the transmitted electric field amplitude (in logarithmic scale) at the upper interface as a function of the bi-layer number ($N$) at $\lambda=1550~nm$. The angle of incidence and the natural logarithm of the $FWHM$ (Full Width at Half Maximum) $\Delta\theta_T$ of the $BSW$ resonance are given on figure \ref{tetaNN}c. As expected, the angular position $\theta_{BSW}$ converges asymptotically but promptly ($N\simeq 7$) to the value corresponding to the pole of the transmission coefficient of the infinite structure. $\Delta\theta_T$ varies exponentially with $N$ (see red with stars line on figure \ref{tetaNN}c) meaning rather similar variations for the $BSW$ resonance quality factor defined by:
\begin{equation}
Q=\cfrac{\theta_{BSW}}{\Delta\theta_T} 
\label{Qequation}
\end{equation}
This exponential increasing of the $Q$-factor with $N$ does not have a real physical meaning because, in practice, losses due to scattering by surface defects and by material absorption lead to a finite value of $N$ (see Fig. \ref{tetaNN}d) for which the structure is optimized (maximum $Q$-factor) \cite{sinibaldi:oex13}. Even if these losses are quite hard to be quantified, most of authors agreed introducing them in calculations by adding a small imaginary part to the optical index.    

Unfortunately, when introducing such absorption \footnote{One notes that absorption is fundamentally proportional to the imaginary part of the permittivity and not to that of the optical index meaning that $\theta_{SBW}$ will be affected.} through $n''$ for all media (except glass substrate), the $BSW$ angular position remains unchanged while the $BSW$ efficiency becomes weaker. In our case, we estimate that $BSW$ excitation is completely canceled for $n''> 10^{-3}$. Figure \ref{tetaNN}d shows the quality factor $Q$ variations versus the number of bi-layers $N$ for different values of the imaginary part $n''$. For loss-less materials, $Q$ tends to infinity as $Q=e^{0.8623N+0.0586}$ while asymptotic behaviors occur for $n''\neq 0$. We have verified that the numerical values appearing in the last relation only depends on the effective index associated to the $BSW$ excitation (here $n_{eff}=n_1\sin{\theta_{BSW}}=1.3928$). 
More importantly, to go further through analytical calculation, it is essential that the transmission coefficient be expressed explicitly as a function of the wave-vector components. Fortunately, in the case of a $BSW$ excitation, the transmission coefficient spectrum can be approached very realistically by a Lorentizan function (see more details in the \textbf{SI} file) leading to express it as:
\begin{equation}
t(k_x,k_y)=\frac{t_{max}}{1+\frac{2i}{\Delta k_x}(k_x-k_x^{BSW})}\cdot 1_{k_y}
\label{lorentz}\end{equation}

Where $k_x^{BSW}$ is the tangential wave-vector component associated to the $BSW$ excitation and $t_{max}$ is the value of the transmission coefficient for $k_x=k_x^{BSW}$ calculated through the $TMM$. 

\underline{\textbf{Modeling of the polarized 3D Gaussian beam}}\newline In the real experiment, a finite beam (commonly Gaussian spatial shape) is used to illuminate the multi-layered structure both in the Krestschmann configuration and either by diffraction. To model such a beam, the plane wave spectrum $PWE$ (or angular spectrum) method is used by coherently summing the amplitude response of all the plane waves composing the Gaussian beam (see \textbf{SI} file for more details). This can be done over the entire structure even inside the layers. The angular spectrum of a 3D polarized Gaussian beams was described in ref. \cite{baida:prb99} and tested several times through comparison with experimental and/or results based on different methods \cite{baida:prb99,baida:ol99,bouhelier:prl03}. An extended formalism from linear to elliptical or circular polarized beam is given by Eqs E.6 -to E.8 of the \textbf{SI} file.

The transmitted electric field distribution associated with the $BSW$ is then calculated in the direct space $Oxyz$ through the equation E.15 of the \textbf{SI} file. The latter involves the transmission Jones matrix of the structure that is basically given by the $TMM$ as $\tilde{t}(k_x,k_y)=-TT_{21}^{-1}\times TT_{22}$ (Eq. E.4 of the \textbf{SI} file). All results calculated through this integral are obtained without any approximation meaning that the vectorial character of both the incident field and the transmission coefficient is taken into account. Nonetheless, due to the resonant character of the transmission, one can reduce the calculation to a scalar equation by only considering the resonant term of the transmission (for instance the $TE$ term in our case) and replacing the transmission coefficient by its expression given by Eq. \ref{lorentz}. 

After fastidious algebra (see \textbf{SI} file), the transmitted electric field amplitude is analytically expressed as a function of the beam-waist $W_0$ and the $FWHM$ ($\Delta k$) of the transmission coefficient through:
\begin{footnotesize}
\begin{eqnarray}
E_t(x,y,z=0)&=&\frac{\sqrt {I_0}t_{max}\Delta k}{4\cos\theta_m}e^{-{\cfrac{8\Delta k\cos\theta_m^2\,x-W_0^2(\Delta k)^2}{16\cos\theta_m^2}}} \notag \\ 
&\times&\left[erf\left(\cfrac{4\cos\theta_m^2\,x-W_0^2\Delta k}{4W_0 \cos\theta_m}\right)
+1 \right] e^{-\cfrac{y^2}{W_0^2}} e^{-ik_x^{BSW} x}
\label{equa19}%
\end{eqnarray}  \end{footnotesize}
Where $erf$ is the error function defined by $erf(x)=\frac{2}{\sqrt{\pi}}\displaystyle \int^x_0{e^{-x^2}} dx$ and $\Delta k=2\pi n_1\cos\theta_m\Delta\theta_T$ is the $FWHM$ of the transmittance spectrum as defined above.

Equation \ref{equa19} provides determining all the $BSW$ properties ($PL$, $L_{GH}$, maximum efficiency...) as it will be discussed below. 

\begin{figure}
\centering
\includegraphics[width=14cm]{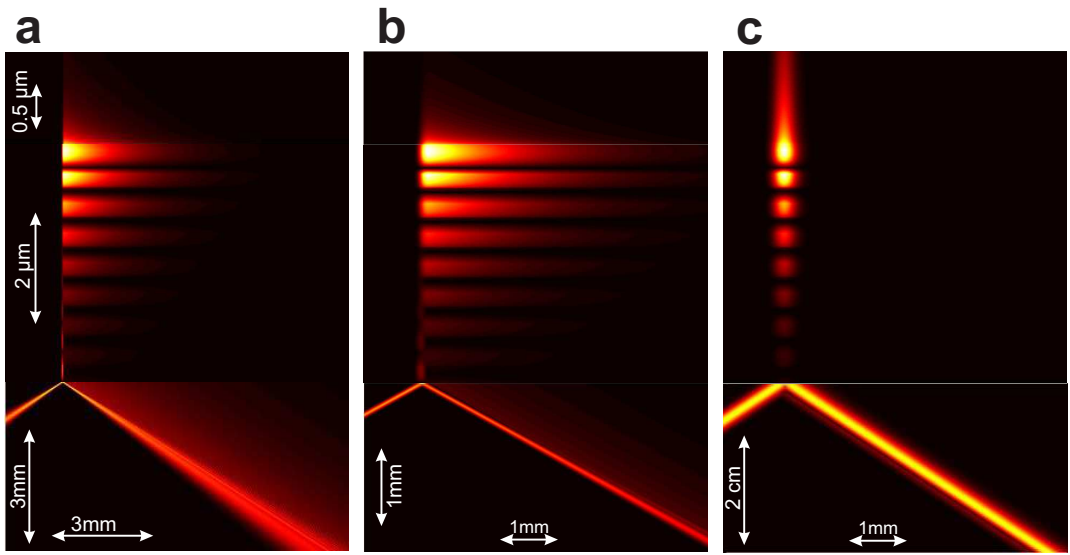}
\caption{\textbf{| Electric field amplitude distributions.} In the three case, the distributions are calculated in a vertical plane passing by the center of the incident beam in the case of $N=7$. The beam-waist of the incident beam is fixed to $W_0=10~\mu m$ in \textbf{a}, $W_0=30~\mu m$ in \textbf{b} and $W_0=1~mm$ in \textbf{c} and it is $TE-$polarized. For the sake of clarity, the electric field was auto-normalized in three different zones: the incidence, the multilayer and the transmission zone. In addition, the scale in the incidence zone varies with the beam dimension as to show both incident and reflected beams.}
\label{imageNN7}
\end{figure}

\underline{\textbf{Results and discussion}}\newline Within $TMM/PWE$ combination (Eq. E.13 of the \textbf{SI} file) one can calculate the electric field distribution over all the structure for any illumination direction, beam-waist or polarization. This versatile character is demonstrated through figures \ref{imageNN7} that present the electric field intensity distribution in the mean plane of incidence ($Oxz$) across all the structure for a $BSW$ excitation within a Gaussian beam of, $W_0=10~\mu m$ in Fig. \ref{tetaNN}a, $W_0=30~\mu m$ in Fig. \ref{tetaNN}b and $W_0=1~mm$ in Fig. \ref{tetaNN}c. The spatial shape of the excited $BSW$ greatly depends on the value of the incident beam-waist $W_0$. When the beam waist is small, the overlapping between the incident beam and the $BSW$ excitation is weak and a small part of the incident energy is coupled to the $BSW$ giving rise to a comet shape for the intensity distribution of the $BSW$ at the top interface. In this case, the determination of the propagation length $PL$ is very easy. When $W_0$ increases (Fig. \ref{imageNN7}b), the angular aperture of the beam decreases and the overlap grows resulting in a more efficient excitation of the $BSW$. Nevertheless, the comet shape becomes less evident due to the competition between the propagation length and the beam width itself. When the beam-waist is very large (Fig. \ref{imageNN7}c), the comet shape completely disappears in the face of the Gaussian shape. In the all three cases, we can clearly see that large electric field confinement occurs in the top layer. For the sake of clarity, the vertical scale in the substrate zone is fixed differently to be large enough to see both the incident and reflected beams. The latter is greatly affected by the $BSW$ excitation and appears to be split into two asymmetric beams when the incident beam waist is small enough due to the presence of out of $BSW$ spectral (angular) components.

From such numerical results on can determine the $BSW$ characteristics corresponding to experimental observed quantities that are recorded on the transmitted near-field, namely the lateral or Goos-H\"{a}nchen shift and the propagation length. Other properties could also be determined such as the Imbert or transverse shift \cite{imbert:prd72}, or the angular shift of the secondary reflected beam \cite{navasquillo:josaa89}. These two last quantities, deriving from the spin-orbit coupling between light and a flat interface, occur on the reflected beam and are mediated by the angular dispersion of the reflection coefficient \cite{bliokh:jo13}. Generally, they need circular or elliptical incident polarization to take place. Furthermore, two different definitions are still used for the $L_{GH}$ assuming it as, the displacement of the maximum of the intensity or, that of the intensity centroid \cite{yallapragada:sr16}. Nonetheless, it is commonly agreed to consider the maximum intensity shift in cases where large propagation distances occur, such as for surface plasmon resonance or $BSW$ \cite{baida:prb99,dubey:eos17}. Consequently, we will restrict our calculation to this last definition as indicating on figure \ref{tetaNN}a. Note that Goos-H\"{a}nchen shift also exists for acoustic waves and was recently studied by analogy with optics \cite{deleo:pageophysics18}. Additional properties dealing with the reflected beam are also reachable as it will be discussed in the following. 

\begin{figure}
\centering
\includegraphics[width=14cm]{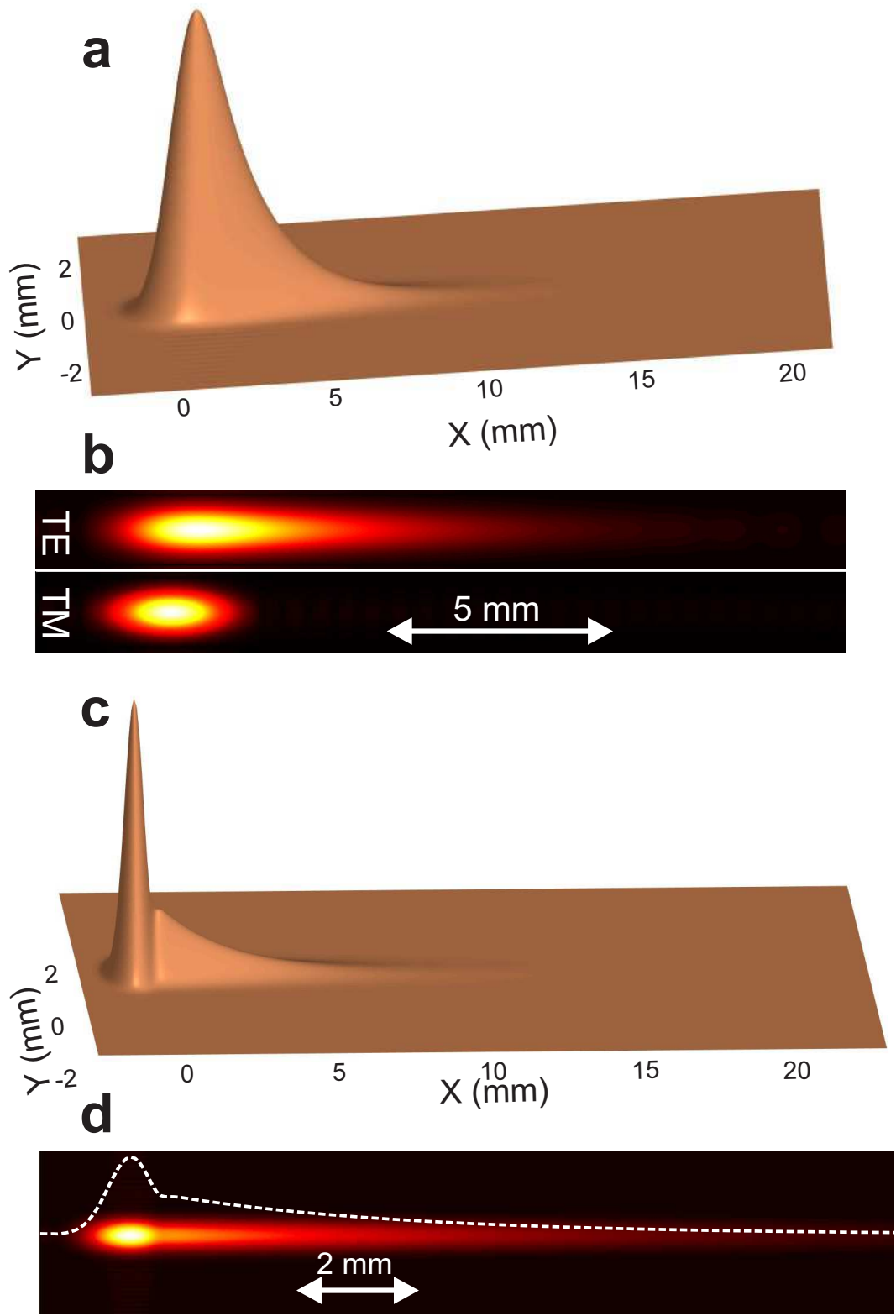}
\caption{\textbf{| 3D map of the electric field intensity distribution.} In all figures, this quantity is calculated at the top interface ($z=0~nm$) for $\theta=\theta_{BSW}$ and $N=7$ and $W_0=300~\mu m$. \textbf{a} and \textbf{c} 3D maps of the electric intensity distribution of the $BSW$. \textbf{b,} Top view maps of the electric intensity for the two polarizations ($TE$ on top and $TM$ on bottom). \textbf{c} and \textbf{d} correspond to an angle of incidence $\theta_m=\theta_{BSW}+1^o$. Experimentally, this kind of distributions is measured by means of scanning near-field optical microscope to estimate both the GH shift (differential value between $TE$ and $TM$) and the propagation length \cite{dubey:eos17}. Note that the intensity maximum value is $70$ times greater in $TE$ than $TM$.}
\label{3D}
\end{figure}

Figure \ref{3D}a shows the 3D map of the $BSW$ electric near-field intensity distribution at $z=0$ from the top interface in a $xOy$ plane as it can be measured by means of Scanning Near-field Optical Microscope ($SNOM$). We can clearly see the surface wave character through the intensity decay that occurs along the propagation distance ($Ox$ here). Top-view distributions are given in Fig. \ref{3D}b allowing identifying the excitation of the $BSW$ (comet shape) in only $TE$ polarization and highlighting the lateral shift that accompanies the excitation of the $BSW$.    

In some recent experimental studies, it was sometimes found that the near-field images of the $BSW$ present a different behavior compared to what is expected (a pure comet shape) as pointed in figure \ref{3D}b. For example, in figure 4b of reference \cite{dubey:eos17} the cross-section made over the intensity map along the propagation direction of the $BSW$ exhibits a depletion next to the maximum. At a first glance, this effect can be attributed to a surface irregularity of the top interface. In fact, by introducing an angular mismatch less than $1^o$ on the angle of incidence, numerical simulations allow reproducing an almost identical behavior as shown in figures \ref{3D}c and d. From figure \ref{3D}a or b, we determine both the spatial position of the intensity maximum that gives $L_{GH}=649~ \mu m$ and the $PL=1.37~mm$. 

\begin{figure}
\centering
\includegraphics[width=14cm]{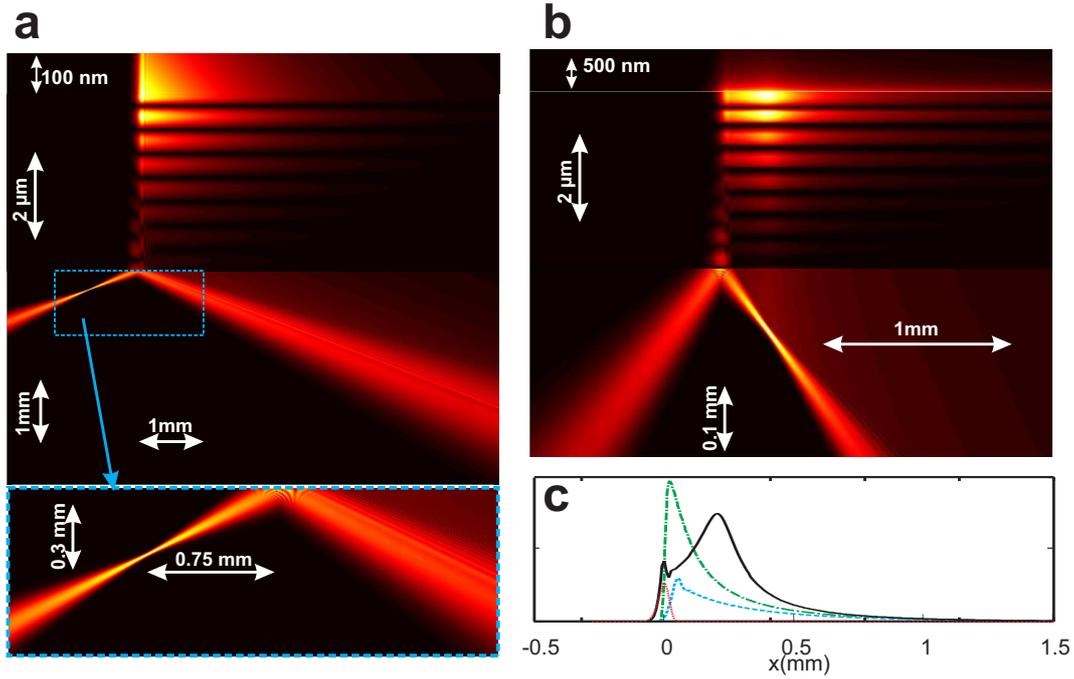}
\caption{\textbf{| Intensity map distributions for defocused Gaussian beams. a,} The waist is at $300~\mu m$ under the first interface and $100~\mu m$ above this interface in \textbf{b.} The beam-waist is set to $W_0=5~\mu m$ and the mean angle of incidence corresponds to the $BSW$ excitation. Dashed blue rectangle in \textbf{a} corresponds to a zoom-in made over the region where incident and reflected beam intersect. The green dashed dotted dotted line on Fig. \textbf{c} corresponds to the electric field intensity along the $Ox$ direction in the case of a waist perfectly centered on the first interface. The red dotted line corresponds to $20\times$ the same quantity in $TM$ polarization. The blue dashed and the black solid lines correspond to \textbf{a} and \textbf{b} cases.}
\label{defocusing}
\end{figure}

Nonetheless, there is another parameter which is difficult to experimentally estimate and which could also affect the excitation of the $BSW$, namely the incident beam defocusing. In fact, in all numerical simulations the beam-waist is supposed to be centered on the top of the substrate. Figure \ref{defocusing} shows two different cases of defocusing. Both of them correspond to the $N=7$-structure illuminated by a Gaussian beam with $W_0=5~\mu m$. The first one (Fig. \ref{defocusing}a) corresponds to a beam-waist located $300~\mu m$ under the PhC structure while it is supposed to be $100~\mu m$ above the substrate-PhC interface in the second (Fig. \ref{defocusing}b). As in Fig. \ref{imageNN7}, the calculated amplitude of the electric field in Fig. \ref{defocusing} is mapped in the $Oxz$ plane with different spatial scales. In the first case, oscillations affect the $BSW$ itself especially near its intensity maximum (see the blue dashed line at the top of the figure (b)) while additional lateral shift of this maximum occurs in the second case (solid black line). This demonstrates how the $BSW$ shape can be affected by a lack of beam focus. In addition, another effect arises on the interference pattern appearing in the reflected beam due to the spatial broadening of the beam falling the PhC. In fact, the total lateral shift at reflection becomes greater and leads to increase the spatial separation between the different angular components of the incident beam. The region where the beam hurts the first interface is emphasized in the blue rectangle in the bottom of Fig. \ref{defocusing}a. One can see the occurrence of curved fringes similar to caustics resulting from the interference between the incident and the reflected angular-wide beams. All this demonstrates the difficulty of interpreting some experimental results but also shows the way to have an effective excitation of the $BSW$.

\begin{figure}
\centering
\includegraphics[width=14cm]{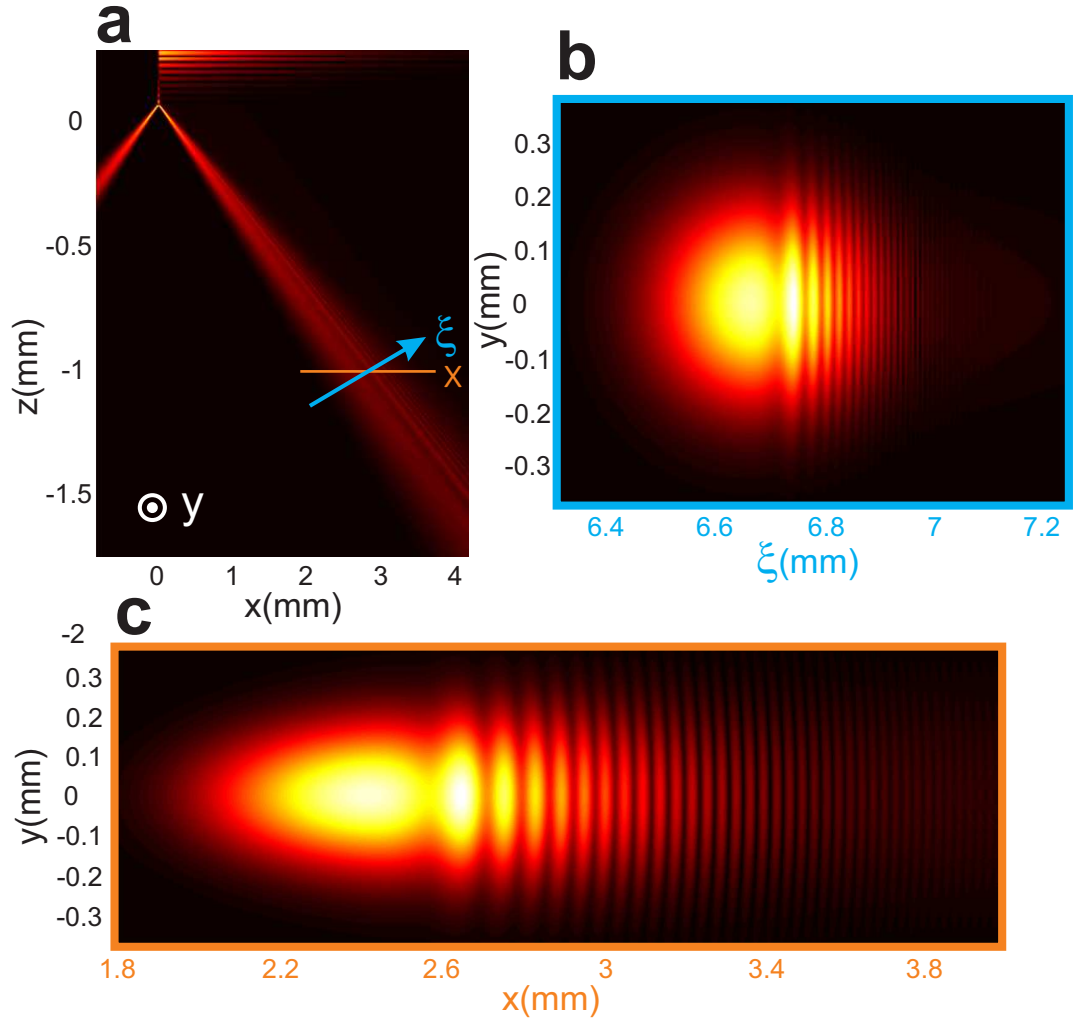}
\caption{\textbf{| Electric field amplitude distributions. a,} In $Oxz$ plane at $y=0$, \textbf{b,} in $xy$ plane at $z=-1~mm$ and \textbf{c,} in $\xi y$ plane that is perpendicular to the propagation direction of the reflected beam (see the $\xi$ axis depicted in \textbf{a}). We recall the geometric parameters: $N=7$, a beam-waist $W_0=5~\mu m$ and a $TE$ polarized incident beam. }
\label{gaussxz_xy_xyz}
\end{figure}

\textbf{The reflected beam}\newline \label{RB} Experimentally, the excitation of $BSW$ is controlled by exploiting the reflected beam (presence of a dip in the reflectance). Consequently, the properties of the latter deserve to be understood to extract information about the $BSW$ excitation. In particular, the oscillation pattern appearing on the reflected beam in the case of strongly focused beams is often highlighted as a signature of the $BSW$ excitation \cite{dubey:as18}. Very recently, Petrova \emph{et al.}\cite{petrova:sr19} exploited the properties of the reflected beam for biosenseng applications. Several theoretical studies have been performed in this context \cite{andaloro:josab05,simon:ol07,konopski:pra12} but all of them considered a 2D-Gaussian beam (prismatic beams) instead of a realistic 3D-beam. In those references, the authors studied the effect of the angular dispersion of the $GH$ shift and they linked it to the fringe pattern that appear on reflection. To point out this phenomenon which occurs also in Surface Plasmon excitation within the Kretschmann configuration, we consider an incident beam with $W_0=5~\mu m$ illuminating the 1D-PhC in the case of $N=7$ and we calculate the electric field distribution in three different planes. Figure \ref{gaussxz_xy_xyz}a shows the electric field amplitude in the $Oxz$ plane as in Fig. \ref{imageNN7}. The fringe pattern is clearly apparent on the reflected beam. A zoom-in over the reflected beam cross-section in the $Oxy$ plane in the substrate, at $z=1~mm$ below the first interface, is shown in Fig. \ref{gaussxz_xy_xyz}b. The spatial oscillations of the electric field intensity are perfectly visible. Although, experimentally, the reflected beam is observed perpendicularly to its propagation direction as in Fig. \ref{gaussxz_xy_xyz}c where the presented electric intensity distribution is evaluated through the $TMM/PWE$ algorithm without any projection operation nor symmetry considerations. According to us, this is the first time that such images are calculated in the case of a real 3D Gaussian beam. In fact, the 2D calculations lead to similar pattern but with different oscillation features. 

\begin{figure}
\centering
\includegraphics[width=14cm]{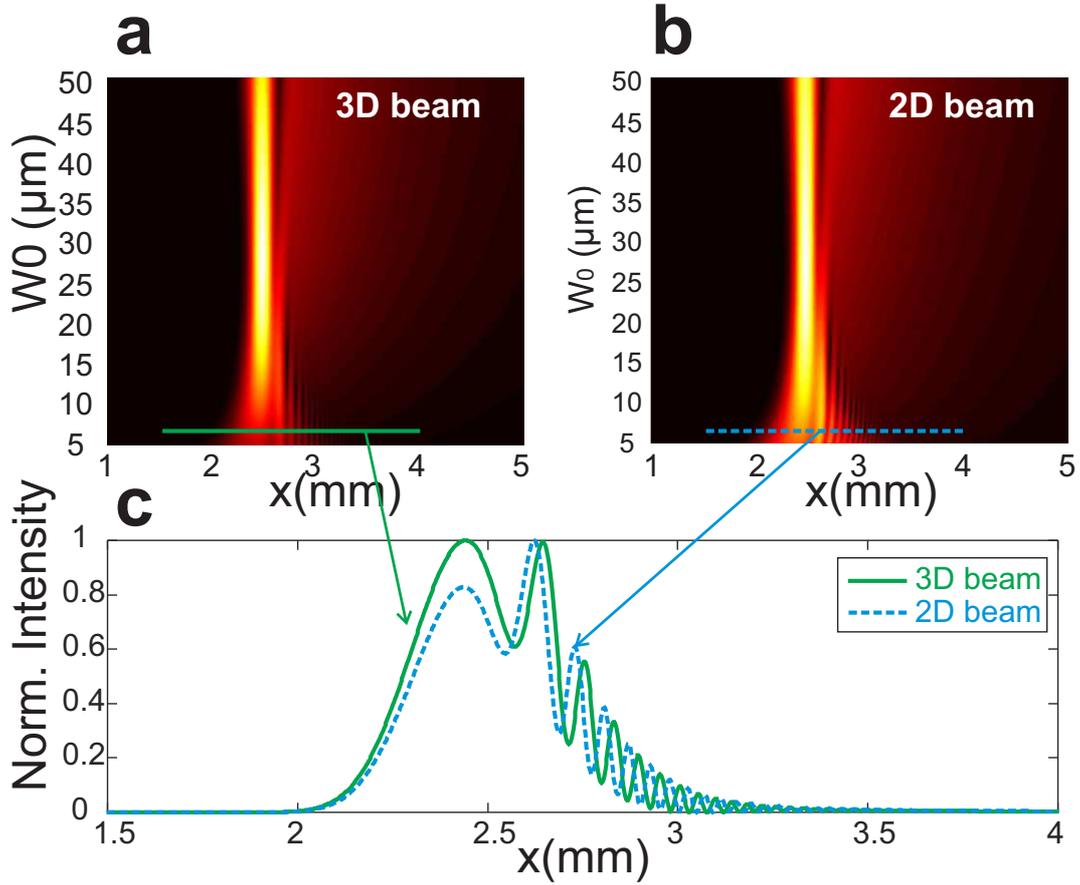}
\caption{\textbf{| Intensity map of the reflected beam} calculated along the $Ox$ axis at $1~mm$ under the PhC as a function of the beam waist $W_0$ when simulations are done for a realistic 3D Gaussian beam in \textbf{a} or a prismatic 2D Gaussian beam \textbf{b}. \textbf{c,} Cross sections made over the two maps for $W_0=6.87~\mu m$ are plotted showing a real discrepancy between the two oscillation patterns.}
\label{oscil_3D_2D}
\end{figure}

Figures \ref{oscil_3D_2D}a and \ref{oscil_3D_2D}b show a transverse cross-section (along the $Ox$ axis) made $1~mm$ under the first interface (substrate-PhC) when the beam-waist varies from $W_0=5~\mu m$ to $W_0=50~\mu m$ for a 3D and 2D Gaussian beams respectively. At a first glance, the two results seem to be very similar. Unfortunately, even if the global shape is comparable, Fig. \ref{oscil_3D_2D}c (where the beam waist was fixed to $W_0=6.87~\mu m$ for both simulations) disclaims it. Although, the oscillations are not at all concordant and their intensity level are clearly different. This is directly due to the contribution of the plane waves that are out of the incidence plane. Indeed, even if the global polarization of the beam is $TE$, these out of incidence-plane waves exhibit $TM$ components whose weight increases as their propagation direction falls out from the plane of incidence. Nonetheless, the Gaussian envelop of the beam amplitude produces a two-lobes amplitude shape (see Fig. 2c of ref. \cite{bouhelier:prl03}) for these components. This is explicitly written in Eqs E.6 of the \textbf{SI} file where the $x-$ and $z-$components of the electric field are not zero even in $TE$ polarization ($\chi=\pi/2 \rightarrow \alpha=0$ and $\beta=1$). The contribution of these $TM$ components to the $BSW$ (in transmission) is negligible because only $TE$ components resonate but this does not prevent their contribution to the reflected beam. 

This interpretation is derived from the Maxwell-Gauss equation ($Div{\vec{E}}=0$) which must be fulfilled for each incident plane wave of the field expansion in the Fourier space. This implies a depolarization term that appears for all waves with wavevector that is not located in the plane of incidence. This is true for both $TE$ and $TM$ polarized beams as shown by Eqs. E.11 and E.12 in the \textbf{SI} file where the field was also expressed in the $TE,TM$ basis. Consequently, the response of a realistic Gaussian beam cannot be calculated by limiting the plane wave expansion over only one spatial frequency component (the $k_x$ one) as it is done in ref. \cite{andaloro:josab05}. In the latter, authors claimed that the $y$-dependence can be suppressed because it does not affect the beam-interface interaction which is rigorously false especially if we deal with the reflected beam. According to us, the same fallacy is at the origin of the clear discrepancy, in terms of oscillation number and amplitude, between the experimental and theoretical results in Fig. 2 of ref. \cite{petrova:sr19}. Therefore, a quantitative exploitation or comparison with experimental results must take into account the contribution of these components.

\textbf{Goos-H\"{a}nchen shift and Propagation distance}\newline\label{LPsection} The number of bi-layers is fixed to $N=7$ in the following as in Fig. \ref{imageNN7} from which one determined the $L_{GH}$ and $PL$ for the three beam-waist values to be: $\{L_{GH}=49.85~\mu m, LP=1.3736~mm\}$ for $W_0=10~\mu m$,  $\{L_{GH}=124.34~\mu m, LP=1.3740~mm\}$ for $W_0=30~\mu m$ and  $\{L_{GH}=1.07~mm, LP=1.3745~mm\}$ for $W_0=1~mm$. Even if the propagation distance is almost constant, its value, in addition to the evolution of the $L_{GH}$, is in clear contradiction with a simple theory based on plane wave analysis \cite{ulrish:josa71,artmann:anph48} estimating these two quantities to be:
\begin{equation}
PL=\cfrac{\lambda}{\pi \Delta\theta_R}\vspace{0.cm}\text{,}\hspace{2cm}
L_{GH}=  \cfrac{-\lambda}{2\pi}\cfrac{\partial \phi}{\partial \theta}\label{lp31}
\end{equation}
where $\Delta\theta_R$ is the $FWHM$ of the dip resonance appearing in the reflectance spectrum and $\phi$ is the phase of the transmission coefficient through the whole structure. Experimentally, the phase variation can hardly be measured. Nevertheless, as it is well-known, this phase value is equal to the half of the reflection coefficient one. Consequently, assuming an interferometric detection (heterodyne), one can reach the reflection phase value. Unfortunately, this proportionality between the two phases of transmission and reflection coefficients is no longer valid when dealing with absorption. However, the $L_{GH}$ of the $BSW$ cannot be obtained by any far-field detection of the reflected beam. Only direct measurement of the near-field allows access to this property.
Theoretically, the variation of $\phi$ versus the angle of incidence is given in figure S3 of the \textbf{SI} file. From this figure, and according to Eq. \ref{lp31}, we estimate the theoretical values of the Goos-H\"{a}nchen shift to be constant ($L_{GH}=770~\mu m$), which cannot be consistent with the calculated values from Fig. \ref{imageNN7} that depend on the beam dimension. This discrepancy needs to be elucidated.

\begin{figure}
\centering
\includegraphics[width=14cm]{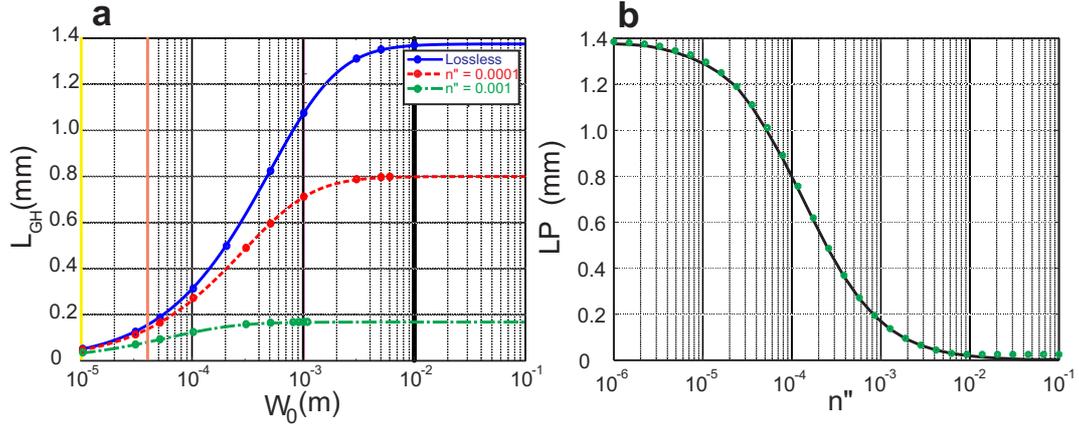}
\caption{\textbf{| Variations of the Goos-H\"{a}nchen lateral shift. a,} in the case of $N=7$ as a function of the beam-waist for three different values of the imaginary part of the optical index of the media (all the layers except the glass and air obviously). The Blue curve corresponds to the loss-less case while red and green ones correspond to $n"=10^{-4}$ and $n"=10^{-3}$ respectively. The yellow, ocher and purple vertical lines correspond the three values of the beam-waist studied in figure \ref{imageNN7}. All the three curves in \textbf{a} correspond to values calculated within the $TMM/PWE$ algorithm while the circles are obtained from Eq. \ref{equa21}. \textbf{b,} Comparison between $TMM/PWE$ results (solid line) and values calculated from Eq. \ref{equa21} (circles) of the $L_{GH}$ as a function of $n"$ in the case of a fixed value of the beam-waist ($W_0=300~\mu m$) .}
\label{GH_NN16_W0ns}
\end{figure}

To this end, we consider the same $N=7$ bi-layer 1D-PhC and we calculate the evolution of the $L_{GH}$ as a function of the beam-waist value through the $TMM/PWE$ algorithm. Figure \ref{GH_NN16_W0ns}a shows that $L_{GH}$ significantly varies with $W_0$ as long as the angular width of the beam ($\frac{\lambda}{\pi W_0}$) is $22$ times larger than the angular width of the $BSW$ (here $\Delta\theta_T=0.642~m rad$) corresponding to $W_0\simeq 1~cm$ as shown by the solid blue line on Fig. \ref{GH_NN16_W0ns}a. The effect of absorption is also studied on the same Fig. \ref{GH_NN16_W0ns}a where we consider the two cases of $n"=10^{-4}$ (red dashed line) and $n"=10^{-3}$ (dashed dotted green line) and study their impact on the $L_{GH}$. As expected, the latter greatly varies with losses. 

Second, the evolution of $L_{GH}$ depicted in figure \ref{GH_NN16_W0ns}a shows an asymptotic behavior limiting it to a maximum value independently from the beam dimension. This behavior is in great contradiction with the results obtained by Konopsky in ref. \cite{konopski:pra12} where formula 2 of that paper states that $L_{GH}$ is proportional to the square root of the beam diameter. Nonetheless, all the demonstrations made in that wonderful paper are formulated for Gaussian beams illuminating an interface near the critical angle in total internal reflection configuration which is different from our case where a sharp resonance with a large $Q-$factor of $1856$ occurs. 

To decide the issue, we make use of the analytical expression of Eq. \ref{equa19}. The spatial position of the transmitted intensity maximum corresponds to the value of $x$ for which the $x-$derivative of the square modulus of the electric field amplitude given by Eq. \ref{equa19} vanishes. This condition leads to Eq. \ref{equa21} that was numerically revolved (see the \textbf{SI} file) for the three cases of Fig. \ref{GH_NN16_W0ns}a. The obtained values are depicted by circles on the same figure showing a perfect agreement with the $TMM/PWE$. We have verified the absolute error to be less than $5\times 10^{-3}$.
 \begin{equation}
\frac{W_0\sqrt{\pi}\Delta k}{4\cos\theta_m}\left[erf\left({\frac{\cos\theta_m L_{GH}}{W_0}-\frac{W_0 \Delta k}{4 \cos\theta_m}}\right)+1\right]-e^{-\left(\frac{\cos\theta_m L_{GH}}{W_0}-\frac{W_0 \Delta k}{4\cos\theta_m}\right)^2}=0\label{equa21}
\end{equation}
For the propagation length ($PL$), Eq. \ref{lp31} cannot be exploited if we assume purely dielectric structure without any absorption because, as mentioned above, the reflectance is equal to $100\%$ and does not exhibit any dip. Nevertheless, introducing a small absorption allows the apparition of a dip in the calculated reflectance. Experimentally, this dip can bring all we need to determine the $BSW$ propagation length $PL$ due to the fact that $\Delta\theta_R\approx\Delta\theta_T$ even in the case of absorption. As determined from Fig. \ref{imageNN7}, we get a propagation length of $PL\approx 1.374~mm$ independently of the beam-waist value while Eq. \ref{lp31} leads to an almost twice smaller value of $PL=769~\mu m$. On notice that expression of $LP$ given by Eq. \ref{lp31} is commonly used to interpret or exploit experimental results \cite{descrovi:oe08,angelini:ol14,dubey:eos17}. Again, we are in front of a contradiction that needs to be clarified.

For this purpose, we will still consider the analytical expression of the transmitted field given by Eq. \ref{equa19} where we can clearly see that for $x\rightarrow\infty (x>>W_0)$, the predominant term is the amplitude expression is $e^{-\frac{\Delta k x}{2}}$ ($erf(\infty)\rightarrow 1$) that corresponds to the electric field behavior far from its maximum. This allowed expressing the propagation length as:
\begin{equation}
LP=\cfrac{2}{\Delta k_x}=\cfrac{\lambda}{\pi n_1 cos\theta_m \Delta\theta_{T,R}}\label{LPequa}
\end{equation}

Replacing $\theta_m = 1.189~rad$, $n_1=1.501$ and $\Delta\theta_T=0.642~mrad$ into Eq.\ref{LPequa} leads to $LP=1.376~mm$ which perfectly agrees the estimated value ($1.374~mm$) by $TMM/PWE$ algorithm.  We have verified the good agreement between the $TMM/PWE$ results (solid line) and this analytical formula (green circles) on figure \ref{GH_NN16_W0ns}b. This perfect agreement between a rigorous numerical method and the mathematical formulation of the transmitted field is an indisputable proof of the accuracy of the two methods. 
\underline{\textbf{Conclusion}}\newline
Combining the $PWE$ with the $TMM$, and using an accurate angular spectrum expansion of a Gaussian beam, turns out to be a powerful tool for simulating and conceiving 1D-PhC structures dedicated to surface wave excitation. The use of the $PWE$ can be extended to integrate any other method ($RCWA$ for instance) able to take into account diffraction by grating (periodic) or by individual pattern such as in \cite{wang:natcom17,descrovi:nl10,yu:sr14,angelini:sr14}. The examples discussed in this paper demonstrate the versatility of this tool that allows highlighting and estimating the unexpected effects of some external parameters (alignment error, or focusing, presence of adhesion layer on the top surface, ...) on the excitation of the surface wave. The major result of this paper is obtained through analytical development that leads to a significant correction of the two important properties of the $BSW$, namely the lateral shift (Eq. \ref{equa19}) and the propagation length (Eq. \ref{LPequa}), for which inaccurate formulas are so far commonly used in the literature. 

\section*{Acknowledgements} Several computations have been performed on the supercomputer facilities of the "M\'esocentre de calcul de Franche-Comt\'e". 
\bibliography{nfo}

\textbf{Competing financial interests}  The authors declare no competing financial interests.
\newpage
\section*{\huge Supplementary Information file}\normalsize
\setstretch{1,4}
\newcommand{\beginsupplement}{%
        
				\setcounter{equation}{0}
        \renewcommand{\theequation}{E.\arabic{equation}}%
        \setcounter{figure}{0}
        \renewcommand{\thefigure}{S\arabic{figure}}%
     }
		
		\beginsupplement
All the numerical simulations were done using codes developed under Matlab environment. These codes combine the $TMM$ and the $PWE$ in Cartesian coordinates. They allows simulations with an arbitrary number of planar membrane stack. The $TMM$ was used alone in the first part of the paper to correctly design a $BSW$ excitation at $\lambda=1550~nm$ as shown on figure 2 of the paper.

\section{The $TMM$ method} 
\begin{figure}
 \begin{center}
\includegraphics[width=11cm]{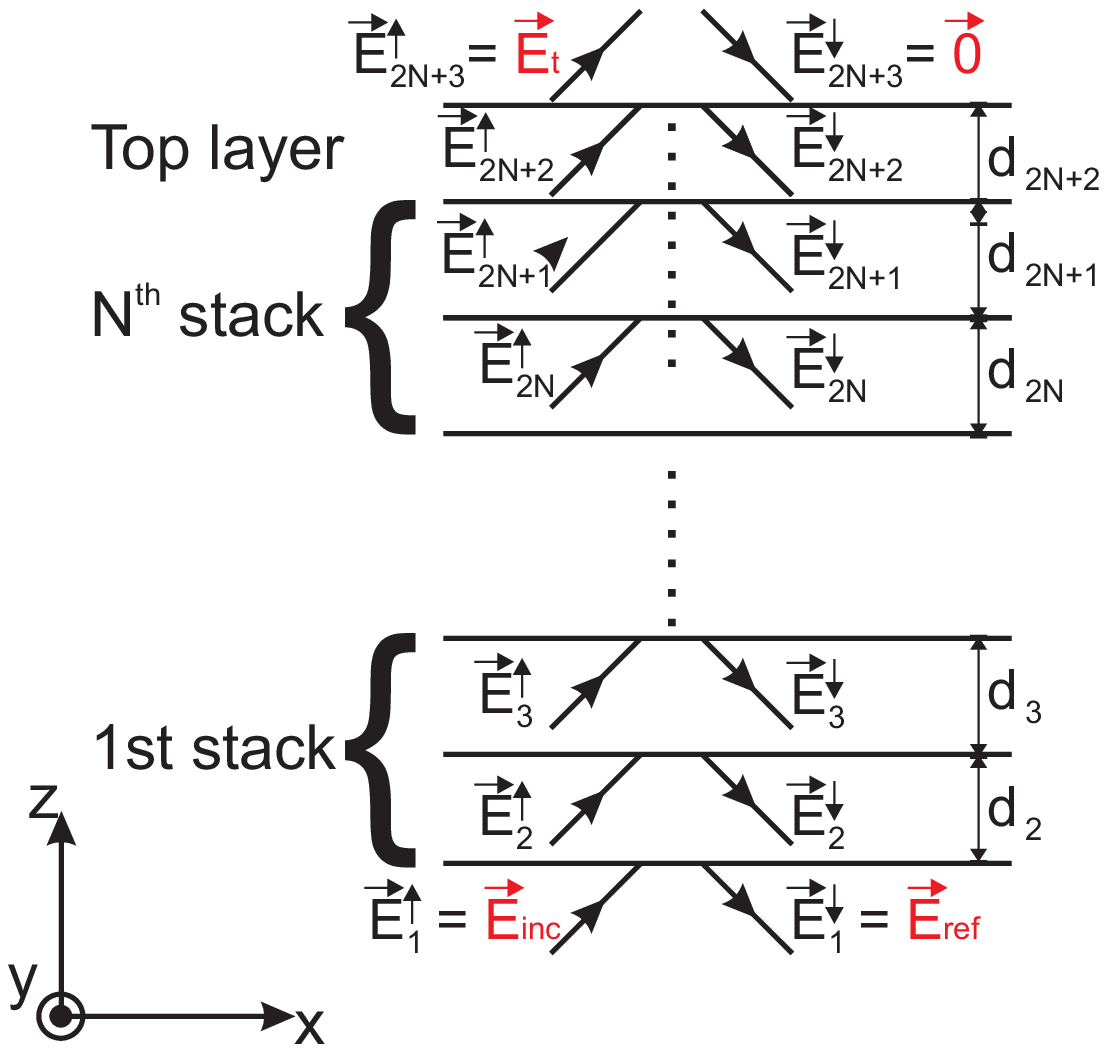}
\caption{\textbf{| Schematic of the stratified structure} built upon $N$ stacks of bilayers and terminated by the top layer where the $BSW$ mode is supposed to take place. The total number of interfaces is then $2N+2$ separating $2N+3$ media including the substrate (media $m=1$) and the superstrate or the transmission media. Each layer is characterized by its dielectric permittivity $\varepsilon_m$ and its thickness $d_m$.}
\label{model} \end{center}
\end{figure}

Let us first recall the principle of the $TMM$. For a given stratified structure of $N$ stacks of bilayers (see schematic of figure \ref{model}), let $M_{m,m+1}$ the transfer matrix linking the electric field components in the medium $m+1$ to those of the medium $m$ in Cartesian coordinates. $M$ is then given by:
\begin{equation}\begin{centering}
\left\{\begin{array}{c}
   E_{m+1~x}^\uparrow \\
   E_{m+1~y}^\uparrow \\
      E_{m+1~z}^\uparrow\\
      E_{m+1~x}^\downarrow\\
      E_{m+1~y}^\downarrow\\
      E_{m+1~z}^\downarrow \\
 \end{array}\right\} =M_{m,m+1}\times\left\{\begin{array}{c}
   E_{m~x}^\uparrow \\
   E_{m~y}^\uparrow \\
      E_{m~z}^\uparrow\\
      E_{m~x}^\downarrow\\
      E_{m~y}^\downarrow\\
      E_{m~z}^\downarrow \\
 \end{array}\right\}\end{centering}
\end{equation}

This is a $6\times 6$ matrix linking the amplitudes in media $m$ (incident $\uparrow$ and reflected $\downarrow$ on the interface separating the two $m$ and $m+1$ media) to the same field components in the $m+1$ media (see figure \ref{model}). In the general case, the matrix dimension is $6\times6$ and cannot be reduced to smaller dimension when we attempt to use it in the case of a Gaussian beam even if the later is $TE$ or $TM$ polarized. This point was discussed in the paper. Nevertheless, it can be reduced to $2\times 2$ matrix in the $TE,TM$ frame (see below).  

In the Cartesian frame, the $M_{m,m+1}$ matrix is given by:\\
$ M_{m,m+1}(k_x,k_y)=$ \begin{tiny}
\begin{equation}
   \left(%
\begin{array}{cccccc}
  \cfrac{w_{m+1}+w_m}{2w_{m+1}}A & 0 & \cfrac{-k_x(\varepsilon_{m+1}-\varepsilon_m)}{2w_{m+1}\varepsilon_{m+1}}A & \cfrac{w_{m+1}-w_m}{2w_{m+1}}B & 0 & \cfrac{-k_x(\varepsilon_{m+1}-\varepsilon_m)}{2w_{m+1}\varepsilon_{m+1}}B \\
  0 & \cfrac{w_{m+1}+w_m}{2w_{m+1}}A & \cfrac{-k_y(\varepsilon_{m+1}-\varepsilon_m)}{2w_{m+1}\varepsilon_{m+1}}A & 0 & \cfrac{w_{m+1}-w_m}{2w_{m+1}}B & \cfrac{-k_y(\varepsilon_{m+1}-\varepsilon_m)}{2w_{m+1}\varepsilon_{m+1}}B\\
  0 & 0 & \cfrac{\varepsilon_mw_{m+1}+\varepsilon_{m+1}w_m}{2w_{m+1}\varepsilon_{m+1}}A & 0 & 0 & \cfrac{\varepsilon_mw_{m+1}-\varepsilon_{m+1}w_m}{2w_{m+1}\varepsilon_{m+1}}B \\
  \cfrac{w_{m+1}-w_m}{2w_{m+1}}A & 0 & \cfrac{k_x(\varepsilon_{m+1}-\varepsilon_m)}{2w_{m+1}\varepsilon_{m+1}}A & \cfrac{w_{m+1}+w_m}{2w_{m+1}}B   & 0 & \cfrac{k_x(\varepsilon_{m+1}-\varepsilon_m)}{2w_{m+1}\varepsilon_{m+1}}B  \\
  0 & \cfrac{w_{m+1}-w_m}{2w_{m+1}}A & \cfrac{k_y(\varepsilon_{m+1}-\varepsilon_m)}{2w_{m+1}\varepsilon_{m+1}}A & 0 & \cfrac{w_{m+1}+w_m}{2w_{m+1}}B & \cfrac{k_y(\varepsilon_{m+1}-\varepsilon_m)}{2w_{m+1}\varepsilon_{m+1}}B \\
  0 & 0 & \cfrac{\varepsilon_mw_{m+1}-\varepsilon_{m+1}w_m}{2w_{m+1}\varepsilon_{m+1}}A & 0 & 0 & \cfrac{\varepsilon_mw_{m+1}+\varepsilon_{m+1}w_m}{2w_{m+1}\varepsilon_{m+1}}B \\
\end{array}%
\right)\notag
\end{equation}\end{tiny}

where $w_m$ is the orthogonal (here $z$) component of the wave-vector given by the dispersion relation $w_m=\sqrt{\varepsilon_m-k_x^2-k_y^2}$, $A$ and $B$ correspond to "upward" and "downward" propagation operator given by $A=e^{iw_md_m}$, $B=e^{-iw_md_m}$ and $d_m$ is the thickness of the layer $m$. Let us emphasize that $w_m$ can be real (propagating waves) or imaginary (evanescent waves), which implies that $B$ is not the complex conjugate of $A$. The total transfer matrix $TT$ is obtained by calculating the matrix product of all the individual matrices respecting the order imposed by the light propagation direction. In the case of $2N+2$ interfaces, its expression is :
\begin{equation}
TT=\prod^{1}_{m=2N+1}M_{m,m+1}=M_{2N+1,2N+2} \times M_{2N,2N+1} \cdots \times M_{2,3} \times M_{1,2}
\end{equation}

Thereby, the electric field in the transmission media is simply expressed as a function of the field in the substrate through:
\begin{equation}\begin{centering}
\left\{\begin{array}{c}
   E_{t~x} \\
   E_{t~y} \\
      E_{t~z}\\
      0\\
      0\\
      0\\
 \end{array}\right\} =TT\times\left\{\begin{array}{c}
   E_{i~x} \\
   E_{i~y}\\
      E_{i~z}\\
      E_{r~x}\\
      E_{r~y}\\
      E_{r~z}\\
 \end{array}\right\}=\left\{\begin{array}{c c}
TT_{11} &TT_{12}\\
TT_{21} &TT_{22}\\
 \end{array}\right\}\times\left\{\begin{array}{c}
   E_{i~x} \\
   E_{i~y}\\
      E_{i~z}\\
      E_{r~x}\\
      E_{r~y}\\
      E_{r~z}\\
 \end{array}\right\}
\end{centering}
\end{equation}
The sub-blocks $TT_{ij}$ are $3 \times 3$ upper triangular matrices meaning that they are invertible. Both reflected and transmitted fields are then expressed in terms of the incident one by:
\begin{eqnarray}
\vec{E}_r(k_x,k_y)&=&-TT_{21}^{-1}\times TT_{22}\vec{E}_i(k_x,k_y) \\
\vec{E}_t(k_x,k_y)&=&\{TT_{11}-TT_{12}\times TT_{21}^{-1}\times TT_{22}\}\vec{E}_i(k_x,k_y) \label{tmatrice}
\end{eqnarray}
The knowledge of the three components of the incident field makes it easy to calculate the transmitted and reflected fields.

\section{{$BSW$ and the other modes of the structure}} 

\begin{figure}
 \begin{center}
\includegraphics[width=14cm]{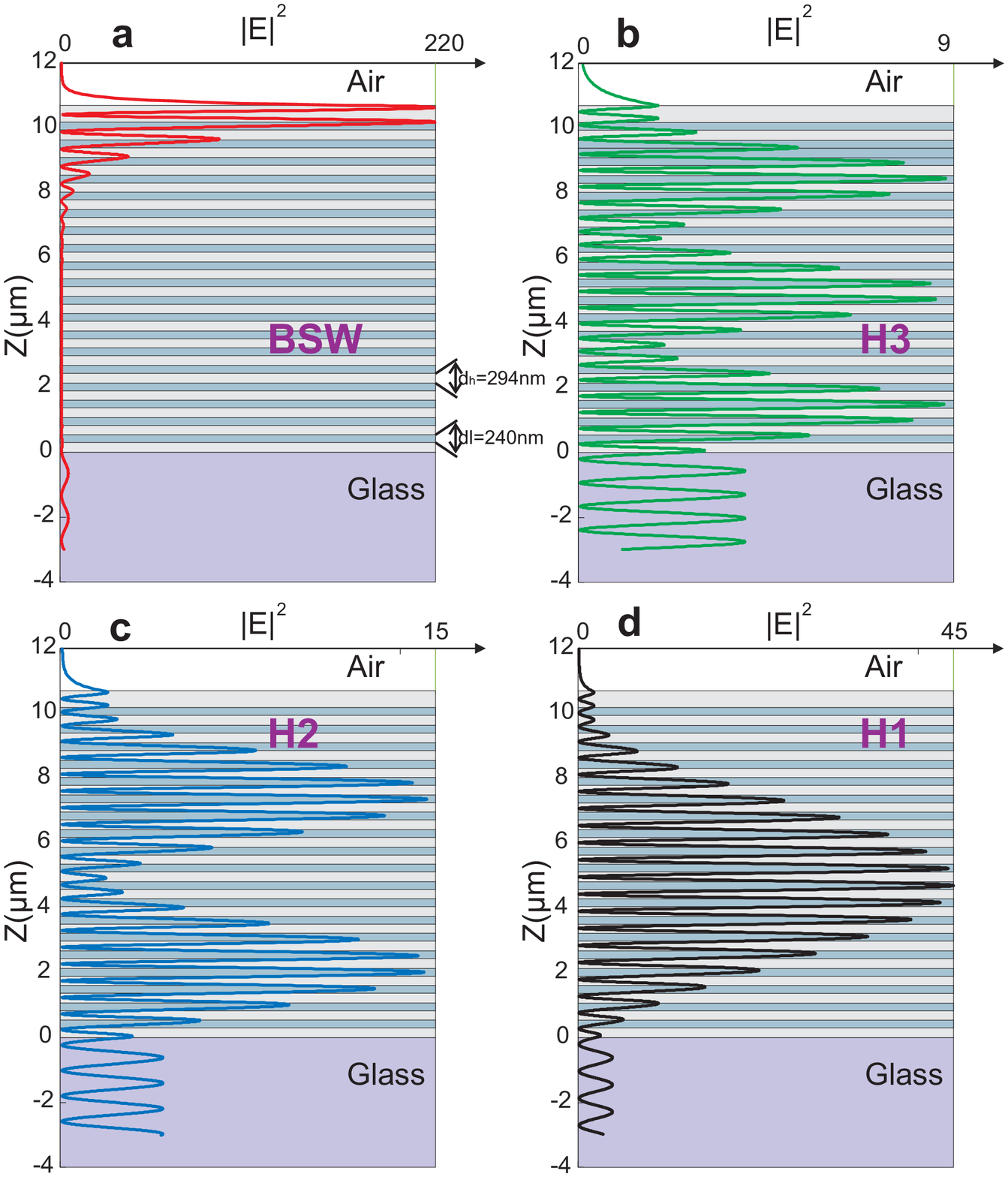}
\caption{\textbf{| Normalized electric field intensity distributions} inside the structure of the four modes of the 1D-PhC structure when $N=20$. \textbf{a,} The $BSW$ excitation, \textbf{b -d,} the three harmonics H1-H3 respectively.}
\label{INTHARM} \end{center}
\end{figure}

As mentioned in the paper and according to figure 1b, supplementary Bloch modes exist at smaller values of the angle of incidence (smaller effective index). They correspond to the cavity modes of the stratified finite structure. Peaks $H_1$ to $H_3$ of figure 1b are the three first harmonics of the total cavity formed by the structure. In addition to the $BSW$ mode presented on figure \ref{INTHARM}a, we show the normalized (to the incident) electric field intensity distribution inside the structure for the three angles of incidence (b-d) in the case of $N=20$-structure illuminated by a plane wave in $TE$ polarization. Light confinement occurs in all cases. As expected, the electric field is localized in the upper layer and leads to large electric field confinement at the upper interface (evanescent wave) when the $BSW$ is excited. This property could be of interest for the enhancement of non-linear effects. 

\section{Plane Wave Expansion of a Gaussian beam}

In addition to the $TMM$, the Plane Wave Expansion $PWE$ (or angular spectrum) is used to take into account the finite size of the illumination (Gaussian beam) through its angular spectrum. Let us recall the plane wave expansion of a polarized Gaussian beam when expressed in the Cartesian frame $Oxyz$ related to the structure ($Oxy$ being the plane parallel to the interfaces as shown in figure \ref{model}):
\begin{footnotesize}\begin{eqnarray} 
E_{x}(k_x,k_y) &=&E_{0}(k_x\cos \theta_m -k_z\sin \theta_m ,k_y)\left\{\alpha-\frac{k_y\sin(\theta_m)}{k_z}\beta\right\} \notag
\\ E_{y}(k_x,k_y) &=&E_{0}(k_x\cos
\theta_m -k_z\sin \theta_m ,k_y)\left\{ \beta\left(\cos\theta_m
+\dfrac{k_x\sin( \theta_m) }{k_z}\right) \right\}  \label{champ}\\
E_{z}(k_x,k_y) &=&-E_{0}(k_x\cos \theta_m
-k_z\sin \theta_m ,k_y)\left\{ \dfrac{k_x\alpha+\beta k_y\cos\theta_m}{%
k_z}\right\} \notag\label{Gausssop} 
\end{eqnarray}\end{footnotesize}
Where $E_{0}(u,v)= \sqrt{\frac{I_0}{2\pi}}W_0\exp \left[-W_0^{2}(u^{2}+v^{2})/4\right]$ is the amplitude of the plane wave characterized by its $(u,v)$ transverse wave-vector components expressed in the proper frame of the Gaussian beam. $k_x$, $k_y$ and $k_z$ are the same components in the $Oxyz$ frame, $I_0$ is the electric intensity of the whole Gaussian beam and $W_0$ its beam-waist defined as the half width at $1/e$ of its maximum amplitude. $\alpha$ and $\beta$ are given by:
\begin{eqnarray}
\alpha&=&cos\chi \label{alpha}
\\ \beta&=&a\sin\chi \label{beta}
\end{eqnarray}
The polarization state of the whole Gaussian beam is defined by the couple ($\chi$, $a$) as:
\begin{itemize}
\item linear directed along the angle $\chi$ measured from the $x$-axis with $a=1$ ($\chi=\pi/2$ for $TE$ and $\chi=0$ for $TM$)
\item circular with $\chi=\pi/4$ and $a=\pm\sqrt{-1}$
\item elliptical with major to minor axes ratio equal to $|\beta/\alpha|=tan\chi$ and $a=\pm\sqrt{-1}$.
\end{itemize} 

Let us emphasize the fact that the combination of the $TMM$ and the $PWE$ can be easily extended to any system of coordinates. Here, all numerical simulations are done in Cartesian coordinates but it is obviously possible to work in the $TE,TM$ frame which is often the case in the literature. Nevertheless, the plane wave expansion of the incident beam must be adapted by electric field projection along these two axis having unit vectors:
\begin{eqnarray}
\vec{e}_{TE}&=&\cfrac{\vec{e}_z\wedge\vec{k}}{\left\|\vec{e}_z\wedge\vec{k}\right\|}=\cfrac{1}{k_{//}}(-k_y,k_x,0)\\
\vec{e}_{TM}&=&\cfrac{\vec{k}\wedge\vec{e}_{TE}}{\left\|\vec{k}\wedge\vec{e}_{TE}\right\|}=\cfrac{1}{k^2}(k_xk_z,k_yk_z,-k_{//}^2)
\end{eqnarray}

where $\vec{e}_z$ is the unit vector along $Oz$ and $k_{//}$ is the tangential (in $xOy$ plane) component of the wave-vector. By projection of the electric field given in Eqs \ref{Gausssop} on these two axes, one gets:\begin{footnotesize}
\begin{eqnarray} 
{E}_{TE}(k_x,k_y)&=&\cfrac{E_{0}(k_x\cos \theta_m -k_z\sin \theta_m ,k_y)}{k_{//}}\notag\\
&\times&\left\{-\alpha k_y+\beta\left(k_x\cos{\theta_m}+\cfrac{k^2_{//}}{k_z}\sin\theta_m\right)\right\} \\
{E}_{TM}(k_x,k_y)&=&\cfrac{k\cdot E_{0}(k_x\cos \theta_m -k_z\sin \theta_m ,k_y)}{k_zk_{//}}\left(\alpha k_x+\beta k_y\cos\theta_m\right)\label{ETETM}
\end{eqnarray}  \end{footnotesize}
As expected, a $TE$-polarized Gaussian beam ($\alpha=0,\beta=1$) exhibits both $TE$ and $TM$ components and likewise for $TM$-polarized beam ($\alpha=1,\beta=0$).
  
In the same basis, the $TMM$ is reduced to a diagonal $2\times 2$ matrix meaning independent relations for $TE$ and $TM$ components. The transmission and reflection coefficients can be than derived from a matricial method \cite{lekner:ndn87} or by applying an algorithm based on the use of Einstein's addition law \cite{vigoureux:josaa92}.  

\section{Analytical expression of the transmitted field} 
In a general case, the three electric field components are calculated by integrating over all the electric fields resulting (in transmission, reflection or inside any layer) from each individual incident plane wave ($k_x,k_y$) through the expression:
\begin{equation}
E_p(x,y,z)=\iint{E_p^R(k_x,k_y)e^{i(k_x x+k_y y+\sqrt{\varepsilon_m\frac{\omega^2}{c^2}-k_x^2-k_y^2}z)}dk_xdk_y}
\label{sop}
\end{equation}
with $p=\{x,y,z\}$ or $\{TE,TM\}$, $\varepsilon_m$ is the dielectric permittivity of media $m$ where the field is calculated, $\omega$ is the angular frequency (pulsation) of the incident light and $c$ is the light velocity in vacuum. The component $E_p^R(k_x,k_y)$ contains the response of the structure to the illumination by the spectral component $(k_x,k_y)$ calculated through the $TMM$ as described previously. It could correspond to the transmitted, reflected or any field inside the structure in medium $m$.

Rigorously, the integral in equation \ref{sop} ranges from $-\infty$ to $+\infty$. Practically, a truncation is numerically introduced by considering only $N_x$ values of $k_x$ and $N_y$ values of $k_y$ in such a way as to correctly describe both the Gaussian shape of the beam and the structure response (resonant modes for instance). Therefore, on the one hand, we consider all plane waves with angular deviation $\delta\theta=\theta_{BSW}-\theta_i\in[-\cfrac{5\lambda}{\pi W_0};+\cfrac{5\lambda}{\pi W_0}]$ which allows taking into account all the angular components of the incident beam with amplitude larger than $\sqrt{I_0}\times 10^{-2}$. On the other hand, the values of $N_x$ and $N_y$ must be as large as possible to prevent any aliasing effect especially if large spatial window are considered. For example, values of $(N_x,N_y)=(6000,350)$ are usually used to calculated the spatial distribution of the electric field over a window of $\Delta x\times \Delta y=150W_0\times 10W_0$. In all cases, convergence tests are systematically applied.  

For the transmitted electric field, $E_p(k_x,k_y)$ is replaced by $t_p(k_x,k_y)E_p^{inc}(k_x,k_y)$ in Eq. \ref{sop}. $t_p(k_x,k_y)$ is the transmission coefficient of the entire structure calculated by the $TMM$ for the incidence given by the two components $k_x,k_y$. In the case of a $BSW$ excitation, the transmission coefficient spectrum exhibits an almost Lorentizan shape (similar to the transfer function of a Band-pass filter) leading to express it as:
\begin{equation}
t(k_x,k_y)=\frac{t_{max}}{1+\frac{2i}{\Delta k_x}(k_x-k_x^{BSW})}
\label{lorentz}\end{equation}

\begin{figure}
 \begin{center}
\includegraphics[width=14cm]{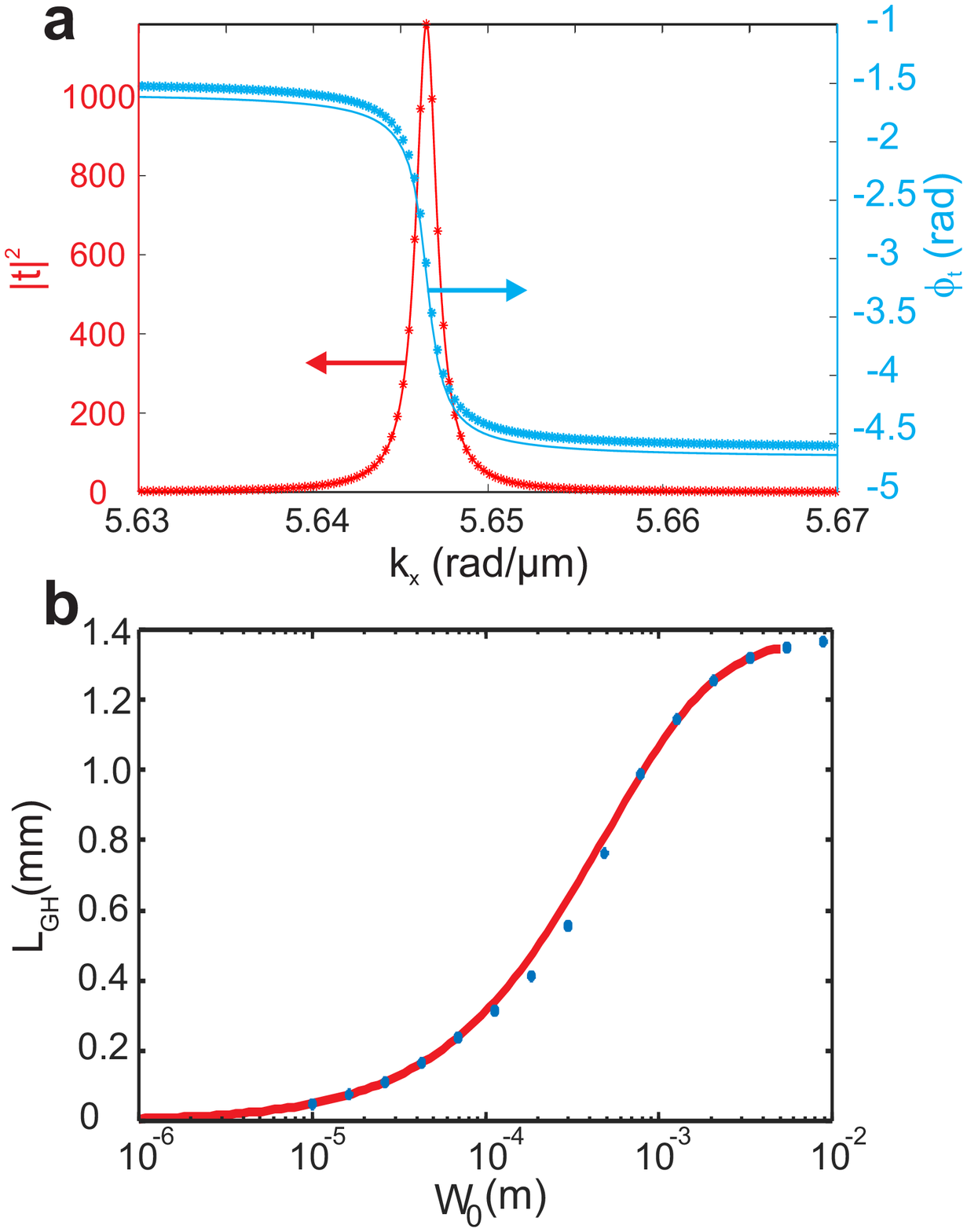}
\caption{\textbf{| a,} Amplitude (red) and phase (blue) of the transmission coefficient of $N=7$-structure in $TE$ polarization. Solid lines correspond to the $TMM$ results while stars are obtained from Eq.\ref{lorentz}. \textbf{b,} The Goos-H\"{a}nchen shift variations as a function of the incident beam-waist $W_0$ in the case of $N=7$ structure. The blue stars correspond to the values calculated by use of the $PWE$ method while the solid red line corresponds to the analytical expression given by Eq. \ref{convol1}.}
\label{compLGH} \end{center}
\end{figure}

Where $k_x^{BSW}$ is the $x-$component of the wave-vector associated to the $BSW$ excitation ($k_x^{BSW}=\frac{2\pi n_1}{\lambda}\sin\theta_m$). Figure \ref{compLGH}a shows the amplitude and phase of the transmission coefficient calculated by the $TMM$ method (solid lines) and by Eq.\ref{lorentz} (stars) in the case of $N=7$ structure. A very good agreement is obtained between $TMM$ and Eq. \ref{lorentz}. We have verified that this analytical model still valid if we increase the $N$ value and also in the case of $BSW$ excited in $TM$ polarization. Expression in Eq.\ref{lorentz} can be extended to more than one Lorentzian resonance by summing the corresponding functions. 

As clearly shown, the transmission coefficient is faithfully described by one complex Lorentizan function. This allows us simplifying the calculation of the integral in Eq. \ref{sop} that becomes:
\begin{equation}
\vec{E}_t(x,y,z)=\iint_{-\infty}^{\infty}{\displaystyle \tilde{t}(k_x,k_y)\cdot\vec{E}_{inc}(k_x,k_y)e^{i(k_x x+k_y y+\sqrt{\varepsilon_m\frac{\omega^2}{c^2}-k_x^2-k_y^2}dk_xdk_y}}
\label{sopt}
\end{equation}
Where $\tilde{t}(k_x,k_y)$ is the $3\times3$ transmission Jones matrix expressed in the $Oxyz$ frame linking the transmission electric field to the incident one. In general, it is given by Eq.\ref{tmatrice} (term between brackets) but in the case of $BSW$excitation, Eq. \ref{sopt} turns into a scalar equation only involving the resonant electric field component (here $TE$ component) and the transmission coefficient given by Eq. \ref{lorentz}.  Let us emphasize that this assumption is only valid for the calculation of the transmitted field and cannot be used for the reflected beam because all the three (in Cartesian) or the two (in $TE,TM$ basis) components of the incident electric field undergo the same reflection coefficient in amplitude (in the case of loss-less materials). 

Accordingly, the transmitted field in the direct space ($Oxyz$) is the convolution product of two Fourier transform functions as follows:
\begin{equation}
\vec{E}_t(x,y,z)=FT\{\tilde{t}(k_x,k_y)\} \otimes FT\{\vec{E}_{inc}(k_x,k_y)\}
\label{convol}\end{equation}  

To go further, we need to express the FT of $t_{TE}$ that is given by:
\begin{equation}\label{tincx}
t_{TE}(x,y,z=z_t)=FT\{t_{TE}\}=\cfrac{2}{\Delta k_x}H(x)e^{-\cfrac{\Delta k_x x}{2}}e^{ik_x^{BSW}x} 
\end{equation} 

For the incident beam, the electric field in the direct space can be expressed through a simple variable change corresponding to a rotation operation (see Eq. 11 of ref. \cite{tamir:josa71}) by:
\begin{equation}\label{Eincx}
E_{TE}(x,y,z=0)=\cfrac{1}{\sqrt{\pi}W_0} e^{\displaystyle -\frac{x^2\cos^2\theta_m}{W_0^2}+\frac{y^2}{W_0^2}+ik_x^{BSW} x}
\end{equation}

By injecting Eqs \ref{tincx} and \ref{Eincx} into Eq. \ref{convol} and after some tricky algebra, we obtain:\begin{footnotesize}
\begin{eqnarray}
E_t(x,y,z=0)&=&\frac{\sqrt {I_0}t_{max}\Delta k}{4\cos\theta_m}e^{-{\cfrac{8\Delta k\cos\theta_m^2\,x-W_0^2(\Delta k)^2}{16\cos\theta_m^2}}} \notag \\ 
&\times&\left[erf\left(\cfrac{4\cos\theta_m^2\,x-W_0^2\Delta k}{4W_0 \cos\theta_m}\right)
+1 \right] e^{-\cfrac{y^2}{W_0^2}} e^{-ik_x^{BSW} x}
\label{convol1}%
\end{eqnarray}  \end{footnotesize}
Where $erf$ is the error function defined by $erf(x)=\frac{2}{\sqrt{\pi}}\displaystyle \int^x_0{e^{-x^2}} dx$.

\section{$LP$ analytical expression} From the last equation, we can clearly see that for $x\rightarrow\infty$, the predominant term is $e^{-\cfrac{\Delta k x}{2}}$ ($erf(\infty)\rightarrow 1$) that corresponds to the electric field behavior far from its maximum. This allowed expressing the propagation length as:
\begin{equation}
LP=\cfrac{2}{\Delta k_x}=\cfrac{\lambda}{\pi n_1 cos\theta_m \Delta\theta_{T,R}}
\end{equation}
where $\Delta\theta_{T,R}$ is the FWHM of $BSW$ resonance peak/dip appearing in the square modulus of the transmission/reflection coefficient spectra.

\section{$L_{GH}$ calculation}
Meanwhile, the Goos-H\"{a}nchen shift corresponds to the value of $x=L_{GH}$ for which the derivative with respect to $x$ of $|E_t(x,y,z)|^2$ vanishes. This equation can be written as:
\begin{equation}
\frac{W_0\sqrt{\pi}}{2 LP \cos\theta_m}\left[erf\left({\frac{\cos\theta_m L_{GH}}{W_0}-\frac{W_0 }{2 LP \cos\theta_m}}\right)+1\right]-e^{-\left(\frac{\cos\theta_m L_{GH}}{W_0}-\frac{W_0 }{2 LP \cos\theta_m}\right)^2}=0\label{LGHequa}
\end{equation}
Unfortunately, this equation cannot be solved analytically. A Root-finding algorithm based on Newton's method is then developed and used to calculate the solution of Eq. \ref{LGHequa}. 
Nevertheless, we verified that the asymptotic value of $L_{GH}$ corresponds to the $LP$ value. This still valid for any configuration with or without absorption as shown on figure \ref{compLGH}b. In the case of $N=7$, the $L_{GH}$ variations versus the incident beam-waist are presented on figure \ref{compLGH}b (solid red line) in comparison with the values obtained through the combined $TMM/PWE$ method (blue circles). We can see a very good agreement between the two curves confirming the extrinsic character of this shift. 

All figures (curves and maps) are obtained within Matlab and presentations are then improved with CorelDRAW. Maple 16.00 version was also used to verify Eq.\ref{convol1}.

\end{document}